\documentclass{aip-cp}

\usepackage[numbers]{natbib}
\usepackage{rotating}
\usepackage{graphicx}
\usepackage[utf8]{inputenc}

\begin{document}

\title{Resonance Extraction from the SAID Analysis}

\author[aff1]{Ron Workman\corref{cor1}}
\eaddress[url]{http://gwdac.phys.gwu.edu}
\author[aff2]{Alfred \v{S}varc}

\affil[aff1]{Institute for Nuclear Studies and Department of Physics \\
The George Washington University \\
725 21st Street NW, 
Washington, D.C. 20052}
\affil[aff2]{Rudjer Bo\v{s}ovi\'{c} Institute, Zagreb, Croatia}
\corresp[cor1]{Corresponding author: rworkman@gwu.edu}

\maketitle

\begin{abstract}
Resonances are extracted from a number of energy-dependent and
single-energy fits to scattering data. The influence of recent,
precise EPECUR data is investigated. Results for the single-energy
fits are derived using the L+P method of analysis and are compared
to those obtained using contour integration applied to the global
energy-dependent fits. 
\end{abstract}

\section{INTRODUCTION}
The SAID website has links to analyses and databases for
a number of fundamental medium-energy reactions. These include
$\pi N$, $NN$, $\pi d$ elastic scattering and the photo- and electro-production
of pions. Fits to data have been either energy-dependent (ED) or single-energy
(SE). In the ED fits, all data are fitted over the full energy
range using a single parameterization. 
In the SE fits, data within narrow energy
bins are fitted by varying the dominant partial-wave amplitudes, with the
ED values taken as a starting point. 
The SE partial-wave analyses (PWA) of $\pi N$ 
elastic scattering have been analyzed for resonance content and have been used
in a number of multi-channel analyses as PWA 'data' in lieu of actual
$\pi N$ scattering data~\cite{Kent,Giessen,BnGa}. 

As the abovementioned multi-channel analyses have found resonances beyond those
extracted from the global ED fits, efforts have been made
to allow for additional resonance 
contributions in the SAID fits~\cite{WI08,SM95,SM90}. 
These we 
review, focusing on the center-of-mass energy region near 1.7 GeV,
where the SAID extractions fail to reliably determine pole positions for the
N(1710)$1/2^-$ and N(1700)$3/2^-$ (PDG 3-star) states~\cite{PDG}.

The SAID SE analyses show more structure than the ED results and recent
Laurent+Pietarinen fits to the SE amplitudes~\cite{LP1,LP2} 
have found poles one could
associate with the abovementioned missing resonances. These results are re-examined
in light of recent, very precise, $\pi N$ elastic scattering data from the
EPECUR collaboration~\cite{EPECUR} 
covering the energy range of interest. These 
measurements were motivated by the search for possible narrow N$^*$ states
near 1.7 GeV, but also serve to distinguish between the existing PWA,
and may also be sensitive to cusp structures associated with opening
channels, such as $K\Sigma$. 
 
Here we should also point out the connection between SAID analyses of 
pion-nucleon elastic
scattering and photoproduction data. As the SAID photoproduction 
analysis~\cite{CM12} is based on the same Chew-Mandelstam formalism used
to analyze pion-nucleon scattering, these reactions share the same
pole and right-hand cut structures. Therefore, by contruction, 
any changes in the resonance
content transfers between the two reactions.

\section{ATTEMPTS TO ADD RESONANCES}

\begin{figure}[h]
  \centerline{\includegraphics[width=450pt]{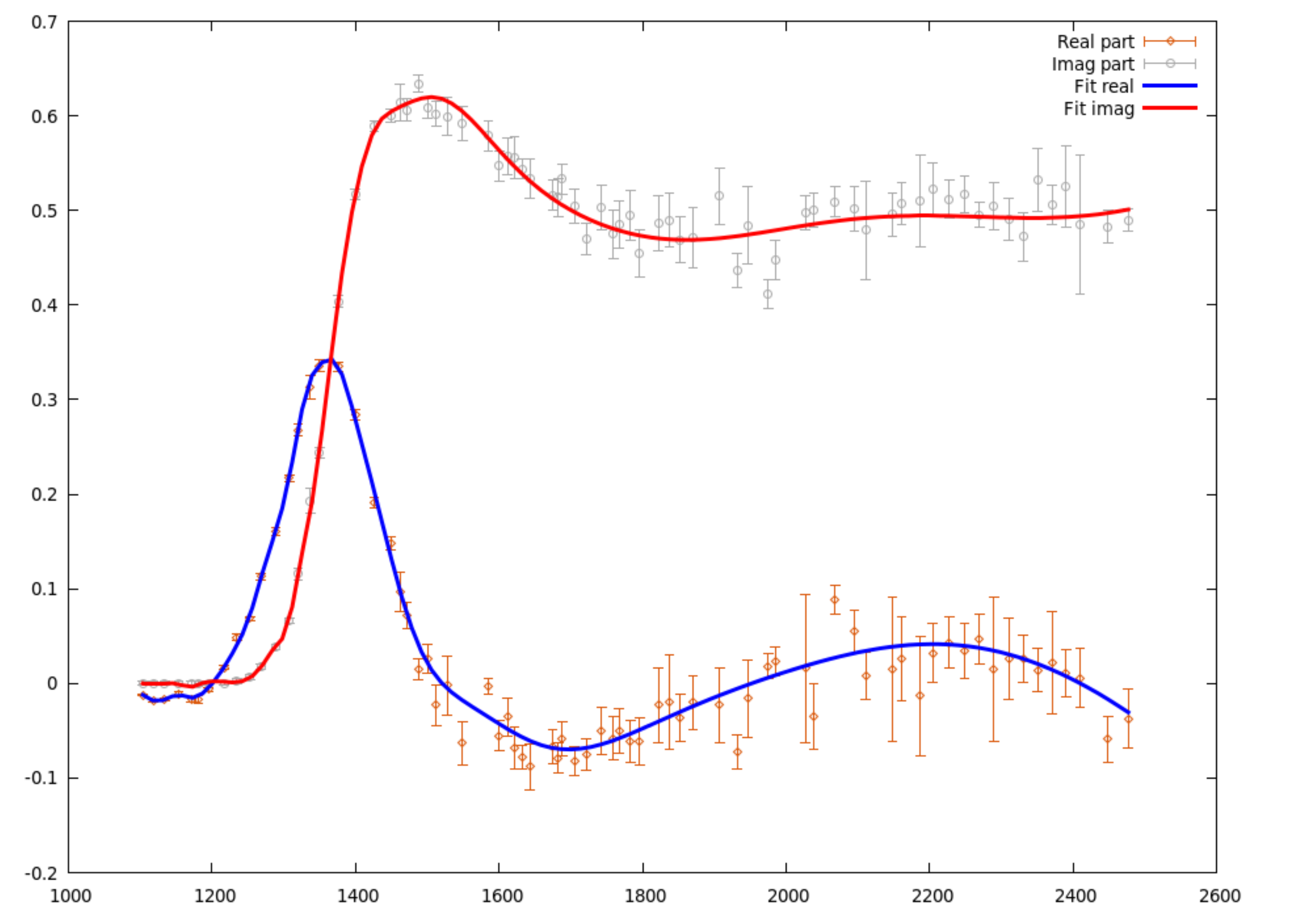}}
  \caption{Fit to SAID SE solutions for the dimensionless 
$P_{11}$ partial wave amplitude, plotted versus center-of-mass energy,
assuming a pole position for the second $P_{11}$ state. See text. Blue (red)
curves are fits to the real (imaginary) parts of the SE amplitudes, represented
as orange (grey) data.}
\end{figure}

\begin{figure}[h]
  \centerline{\includegraphics[width=450pt]{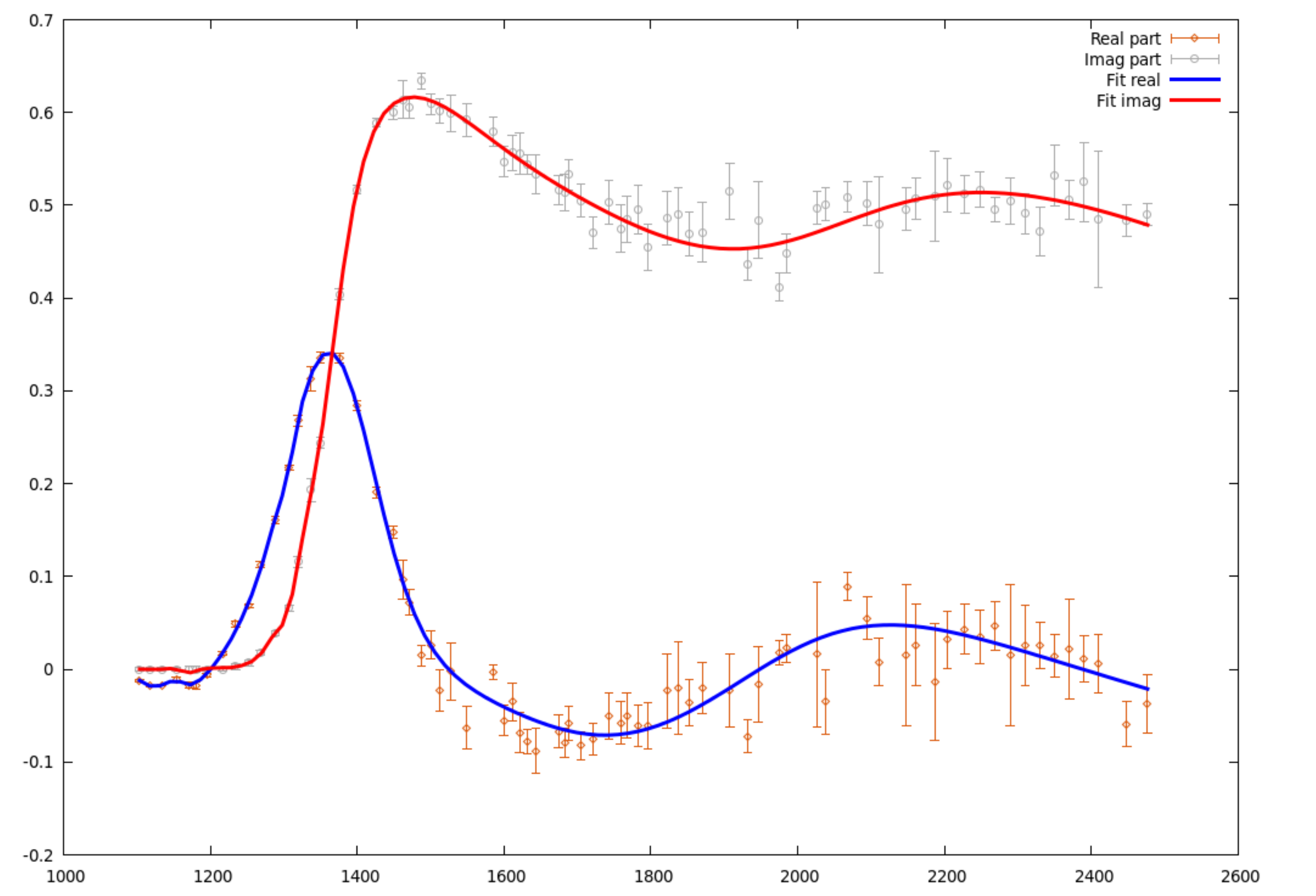}}
  \caption{Fit to SAID SE solutions for the dimensionless 
$P_{11}$ partial wave amplitude, plotted versus center-of-mass energy,
assuming a fixed second pole position, and fitting a third pole position.
Notation as in Fig.~1.}
\end{figure}

\begin{figure}
  \centerline{\includegraphics[width=450pt]{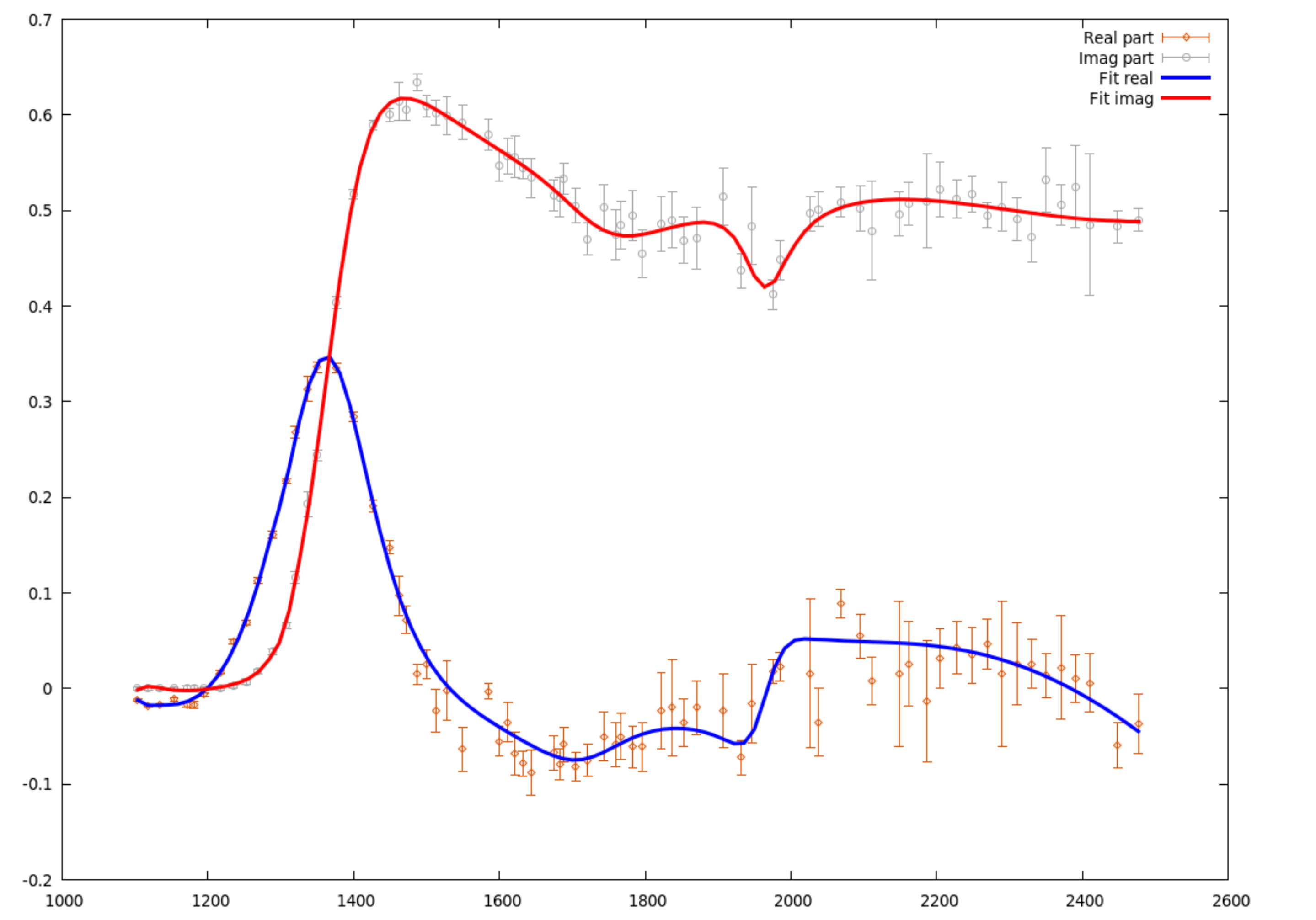}}
  \caption{Fit to SAID SE solutions for the dimensionless $P_{11}$ partial wave amplitude, plotted versus center-of-mass energy,
with three fitted pole positions. Notation as in Fig.~1.}
\end{figure}
 
\begin{table}[h]
\caption{Summary of N(1710) pole determinations from the KH~\cite{KH80}, CMB~\cite{CMB}, BnGa~\cite{BnGa} and the L+P analyses: L+P(WI08), L+P(EPECUR), and L+P(Bin Shift) . }
\label{tab:a}
\tabcolsep7pt\begin{tabular}{lll}
\hline
\tch{1}{l}{b}{Fit}  & \tch{1}{l}{b}{Real}  & \tch{1}{l}{b}{-2 Imaginary}  \\
\hline
KH & 1690 & 200\\
CMB & 1690$\pm$20 & 80$\pm$20 \\
BnGa & 1687$\pm$17 & 200$\pm$25 \\
L+P(WI08) & 1711$\pm$10$\pm$0.6 & 84$\pm$20$\pm$2 \\
L+P(EPECUR) & 1725$\pm$18 &200$\pm$37 \\
L+P(Bin Shift) & 1653$\pm$22 & 168$\pm$41\\
\hline
\end{tabular}
\end{table}

As resonances, apart from the $\Delta(1232)$, have appeared 
in the SAID analysis
as a result of the Chew-Mandelstam formalism, and are not inserted by hand,
only those states with significant $\pi N$ couplings have been found,
constituting a 'minimal' set of resonances. However, the classic 
Karlsruhe-Helsinki~\cite{KH80} (KH) and 
Carnegie-Mellon-Berkeley~\cite{CMB}(CMB) analyses have
found numerous additional (PDG 3-star and lower rated) resonances, with some
of these now appearing more clearly in reactions with different final states
(such as $K\Lambda$)~\cite{BnGa}. 

In the 1990 SAID analysis~\cite{SM90}, amplitudes from the KH and CMB 
fits were added as soft constraints, and fitted together with the scattering
data, in order to force the SAID fit to more closely approximate the KH and CMB results. A
number of new poles appeared in these constrained SAID analyses, but these were generally
not in good agreement with the KH and CMB values. In particular, the 
$P_{11}$ and $D_{13}$ waves failed to produce poles near the expected
$P_{11}$(1710) and  $D_{13}$(1700) states, absent from the original SAID analysis.

A sensitivity test for resonance addition was made in the 
1995 analysis~\cite{SM95}. Here the standard fit was augmented in a product
S-matrix approach, $S_{\rm Prior} S_{BW}$, with a Breit-Wigner state added
and searched in each partial wave. This exercise found evidence for a 
second $P_{11}$ state somewhat higher in mass and broader than expected.
In addition, a second $F_{15}$, N(1860)$5/2^+$, was found (a PDG 2-star
state) which persists in current fits. Combining this search with the
previous set of soft amplitude constraints was no longer feasible as 
amplitude constraints were being used in the iteration of fits constrained
by forward and fixed-t dispersion relations.

Finally, a fit was made with explicit Chew-Mandelstam K-matrix poles 
inserted in each partial wave~\cite{WI08}. The initial expectation was
for a significant increase in the number of T-matrix poles per partial
wave. However, the actual result was a set of partial wave amplitudes
nearly identical to those from a fit without explicit K-matrix poles. In
effect, the fit generally moved the dominate structures to the poles
added by hand, with secondary poles mainly appearing far from the physical
axis. For example, the second $P_{11}$ pole appeared at (1646 - i290) MeV
with a real part near the expected value but double the expected imaginary
part. In the $D_{13}$ wave, two poles appeared again with the expected real
part but with imaginary parts that were either too small (accompanied by
a very small residue) or too large, compared to values found in the KH
and CMB fits.

\section{LAURENT+PIETARINEN FITS}

While it has proven challenging to incorporate additional pole structures
into the SAID ED fits, the associated SE fits have, by design, added 
structures in energy beyond the global ED results. By analyzing narrow
energy bins of data, without any smoothness constraints apart from the
ED amplitude starting points, the SE amplitudes would be expected to 
fluctuate around the ED values. The original intent of these SE fits
was to check for systematic structure missing from the ED parameterization.
However, 
in order to determine whether the SE behavior is consistent with additional
pole structure, it must first be fitted with a function 
that can be extrapolated into the complex energy plane. A convenient form
is the Laurent+Pietarinen (L+P) parameterization of the $\pi N$ T-matrix,
\begin{equation}
T(W) = \sum_{i=1}^{k} {{a_{-1}^{(i)} } \over {W - W_i}} + B(W),
\end{equation} 
where the non-pole term $B(W)$ is constructed from a conformal mapping of the
cut energy plane onto the unit circle as described in Refs.~\cite{LP1,LP2}.

Fits of the L+P type have~\cite{LP1,LP2} have found, for the $P_{11}$ 
partial wave, the first (Roper) state and the second N(1710) with pole
values in line with PDG estimates, starting from the WI08 SE amplitudes~\cite{WI08}.
For the second $D_{13}$ state, the L+P fit finds a pole at 
(1752 - i286) MeV, qualitatively in line with the deeper pole found in the 
explicit pole fit of Ref.~\cite{WI08} but outside the, rather broad,
PDG imaginary-part range of approximately 50-200 MeV. 

One might ask if the SE fluctuations are reliable or are due to possibly
inconsistent data in particular energy bins, or are influenced by the choice
of energy-bin width and center. We address both questions below by first
considering the effect of including much more precise cross section data
available from the EPECUR experiment~\cite{EPECUR}, and then shifting the bin 
positions and widths.

\section{MODIFIED ED AND SE FITS}

As a first step, the new EPECUR data were included in a revised ED fit to
update to the WI08 ED result. This ED solution was then used as the
basis for new SE solutions. As the EPECUR data are very precise and have
a small step size in energy, individual energy bins contained significantly
more high-quality cross section data. This had the effect of reducing 
the SE errors by up to approximately 30 percent, depending on the partial
wave. 

The SE set for the $P_{11}$ partial wave was then fitted using a number of
assumptions regarding the pole content, as displayed in Figs.~1 to 3. In
Fig.~1, a 2-pole search was made with the second fixed at the value found
in the corresponding ED fit, (1659 - i262) MeV. The resulting chi-squared
was 1.1 per datum for the L+P fit. Adding a third pole, with the second
remaining fixed, as shown in Fig.~2, had little effect on the qualitative
fit and chi-squared. However, allowing a search of the second pole resulted
in a qualitatively different fit with a slightly better chi-squared 
per datum. The resulting second pole position was (1725 - i100) MeV, which
is in good agreement with the present PDG estimate of (1720 - i115) MeV. 

As a final exercise, the SE energy bin centers and widths were changed
randomly to determine whether this would produce significantly different
scatter in the SE solutions. The resulting SE set was again fitted with
three searched poles. In this case, the second pole, which one would
expect to correspond to the N(1710), was found at (1653 - i84) MeV, 
significantly different from the previous determination. This then 
suggests re-binning effects should be taken into account when estimating
extracted pole position uncertainties. Results for the N(1710) pole 
determinations are summarized in Table~1. 
  
\section{ACKNOWLEDGMENTS}
This work was supported in part by funds from the George Washington
University and by the U.S. Department of Energy (Offices of Science and
Nuclear Physics, Award No. DE-FG02-99-ER41110 ).

\nocite{*}
\bibliographystyle{aipnum-cp}%
\bibliography{workman}%

\end{document}